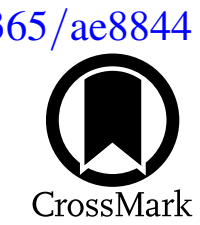

# Statistical Analysis of Minifilament Eruptions Using Full-disk Hα Blue-wing Observations at Big Bear Solar Observatory

Artin Khaleghi[1], Qin Li[1,2], Nengyi Huang[1,2], Jeongwoo Lee[1,2], and Haimin Wang[1,2]
[1] Institute for Space Weather Sciences, New Jersey Institute of Technology, University Heights, Newark, NJ 07102-1982, USA; ak3275@njit.edu
[2] Center for Solar-Terrestrial Research, New Jersey Institute of Technology, University Heights, Newark, NJ 07102-1982, USA

## Abstract

This paper presents a comprehensive statistical analysis of minifilament eruptions (MFEs) using high-cadence full-disk Hα blue-wing observations from Big Bear Solar Observatory. Despite the recognized importance of MFEs for coronal dynamics and solar wind structuring, previous efforts were often limited by small sample sizes or restricted fields of view, leaving open questions about their global occurrence and characteristic properties. We developed an algorithm incorporating intensity thresholding and temporal tracking to detect sudden enhancements in Hα blue-wing images. A total of 1986 such events were identified during a 4 hr period on 2020 June 9. These detections were cross-validated using Hα line-center observations to confirm the presence of associated filament structures. The analysis yields an occurrence rate of $\sim 6.6 \times 10^4$ per day, an average length of ∼17 Mm, and a typical lifetime of ∼21 minutes. A power-law distribution in eruption lengths (with slope $\sim -4.8$) and a sublinear scaling between duration and length suggest that larger eruptions tend to last longer. The length distribution of MFEs was further analyzed relative to active region proximity, revealing that eruptions occurring nearby a sunspot region were, on average, larger than the global mean eruption length. Additionally, an eruption density map was used to analyze the spatial distribution of MFEs relative to coronal hole boundaries. The results indicate a mild suppression of activity inside coronal holes and intermediate eruption frequency in the boundary regions. We briefly discuss their potential role in structuring the small-scale solar wind features such as magnetic switchbacks and small-scale magnetic flux ropes.

## 1. Introduction

Filaments are elongated magnetic flux-rope structures in the solar atmosphere, composed of relatively cool and dense plasma suspended by magnetic fields above the photosphere (H. Zirin & E. Tandberg-Hanssen 1960; D. M. Rust & A. Kumar 1994; E. Tandberg-Hanssen 1995). Minifilaments are their smaller-scale counterparts, sharing similar morphology and magnetic configurations but occurring on shorter spatial and temporal scales (J. Wang et al. 2000). The eruption of a minifilament (MFE) involves magnetic reconnection, during which stored magnetic energy is released. They are often observed to be associated with localized brightenings and may be related to the formation of small-scale coronal jets (A. C. Sterling et al. 2023). During their frequent occurrence, MFEs may play an important role as a small-scale dynamic phenomenon in the solar chromosphere. MFEs could be the potential source for coronal heating, mass transfer between solar winds and the solar corona, and ultimately for space weather variability (N. Huang et al. 2023).

MFEs are frequently observed together with coronal jets, suggesting that an MFE can trigger a jet through reconnection with surrounding magnetic fields (A. C. Sterling et al. 2016). This eruptive process mirrors that of large-scale filament eruptions but on a smaller scale, both in energy and duration. Observational studies using high-resolution instruments such as Hinode and Solar Dynamic Observatory (SDO) have repeatedly confirmed that many coronal jets are preceded by, or coincide with, the eruption of small-scale filaments embedded in magnetic bipoles at coronal hole (CH) boundaries or in quiet-Sun regions (R. L. Moore et al. 2010; C. J. Schrijver et al. 2011). Observationally, many coronal jets are triggered when a destabilizing minifilament erupts and reconnects externally with ambient open field, launching narrow jet spires (P. Kumar et al. 2019; A. C. Sterling & R. L. Moore 2020). In some cases, MFEs have been shown to lead to blowout jets, where both cool filament material and hot plasma are expelled simultaneously (R. L. Moore et al. 2010). Further, erupting minifilaments can re-form at the same neutral line and trigger multiple jet events sequentially. For example, B. Yang et al. (2019) observed a minifilament that partially and then fully erupted on the same neutral line, driving three successive two-sided loop jets from the same location. Similarly, A. C. Sterling et al. (2016) analyzed jet-producing active region (AR) events and found that multiple EUV/X-ray jets were driven by successive eruptions of small (≈20″) minifilaments from the same region—each preceded by flux cancellation and followed by jet bright-point activity. The recurrence of MFEs in jet-producing regions highlights their role as a fundamental building block of small-scale solar activity and their relevance to the mass and energy balance of the solar atmosphere.

In the magnetic-breakout picture, eruption of a minifilament destabilizes the overlying arcade and reconnects with surrounding open field, launching narrow jet outflows and—when both cool filament material and hot plasma escape—blowout jets (P. F. Wyper et al. 2018; Y. Shen et al. 2019). Recent observational and modeling work suggests that such small-scale reconnection and flux-rope formation at the Sun can produce magnetic structures that are measurable in the heliosphere. In particular, Parker Solar Probe (PSP) observations and theory indicate that sharp deflections in the near-Sun magnetic field known as "switchbacks" can originate from interchange or bursty reconnection, producing small flux ropes and locally twisted field lines close to the Sun (J. F. Drake et al. 2021; R. Laker et al. 2021). A broad body of in situ studies has recorded numerous small-scale magnetic flux ropes in the solar wind and linked their topology, occurrence rates, and signatures to coronal reconnection processes and to local generation in evolving solar-wind shears (Q. Hu et al. 2004; M. L. Cartwright & M. B. Moldwin 2010; J. Zheng & Q. Hu 2016). Taken together, these results support a picture in which MFEs and associated breakout/interchange reconnection may not only power coronal jets but also contribute to the population of small, erupting flux-rope structures that seed fine-scale magnetic variability in the emerging solar wind (N. Huang et al. 2023; H. Farooki et al. 2024).

The observation of filament eruptions historically utilized multiple chromospheric spectral lines, such as H$\alpha$, Ca II K, and H$\beta$ (H. Zirin & E. Tandberg-Hanssen 1960). Among these, H$\alpha$ is the most widely used due to its strong contrast, where filaments appear in absorption against the solar disk, as well as its wide availability from ground-based observatories (J. Wang et al. 2000; M. M. Kuraica et al. 2009; I. Ermolli et al. 2010; V. Capparelli et al. 2017). H$\alpha$ observations, particularly in the line center and Doppler wings, provide valuable diagnostics for detecting filaments and tracking their dynamic evolution (J. Wang et al. 2000; M. M. Kuraica et al. 2009). Enhanced dark features in the H$\alpha$ blue wing often correspond to rising material, as Doppler blueshifts shift the absorption profiles toward shorter wavelengths. This makes the blue wing a suitable channel for identifying upward motion associated with MFEs (M. M. Kuraica et al. 2009). However, in the absence of red wing observations, the interpretation of such blue-wing absorption must be treated with caution, as symmetric line broadening caused by localized heating or turbulence can also enhance wing absorption without implying bulk motion. Despite this limitation, our study provides one of the first statistical assessments of the occurrence rate of MFEs based on blue-wing signatures. Although high-resolution H$\alpha$ observations in multi-wavelengths across both blue and red wings have been available from ground-based observatories that have a limited field of view, the global-scale observations are typically in H$\alpha$ line center. During BBSO's support of the PSP mission, the full-disk observations are tuned to the far blue wing of H$\alpha$, facilitating the detection of higher-speed blueshifted structures. Many of those structures are likely MFEs.

Among other factors, the spatial distribution of MFEs is essential for understanding the magnetic and plasma environments for small-scale solar activity. ARs are magnetically concentrated areas on the solar surface, often associated with sunspots and strong, evolving magnetic fields. They are primary sources of solar eruptions, such as flares, coronal mass ejections (CMEs), and filament eruptions (L. van Driel-Gesztelyi & L. M. Green 2015; I. Kontogiannis et al. 2024). Previous studies have highlighted that MFEs frequently occur near ARs, particularly along their edges, where dynamic magnetic activity such as flux emergence and cancellation are prevalent (A. C. Sterling et al. 2015; N. K. Panesar et al. 2018). These conditions are conducive to reconnection-driven eruptions, including minifilaments and jets. While statistical examinations of such spatial relationships remain limited, case studies suggest that MFEs at or near solar ARs tend to be larger than their quiet-Sun counterparts. For instance, A. C. Sterling et al. (2016) noted that the AR minifilaments in their study were noticeably larger than the quiet-Sun minifilaments previously observed in polar coronal holes (A. C. Sterling et al. 2015). Investigating the size distribution of MFEs relative to ARs offers insights into the role of local magnetic topology in triggering small-scale eruptions and may help to distinguish between AR-driven and quiet-Sun minifilament dynamics.

In contrast, coronal holes are regions of the solar corona characterized by lower plasma density and temperature, and appear darker in extreme ultraviolet (EUV) and X-ray observations. These regions are magnetically open, allowing charged particles to escape into the heliosphere and form high-speed solar-wind streams (S. R. Cranmer 2009). Coronal holes are of particular interest because they serve as one of the primary sources of fast solar wind and have a significant impact on space weather conditions near Earth (S. Cranmer 2002; T. H. Zurbuchen & I. G. Richardson 2006). The open magnetic field lines in these regions facilitate the outflow of plasma, contributing to geomagnetic disturbances when these high-speed streams interact with Earth's magnetosphere (S. J. Hofmeister et al. 2019). Furthermore, the magnetic and plasma conditions within coronal holes are not typically conducive to magnetic reconnection processes that drive eruptive phenomena, such as MFEs. Although coronal holes typically lack the magnetic complexity required for reconnection-driven eruptions, their boundaries can host interchange reconnection between open and closed field regions to produce small-scale outflows, coronal jets, and even MFEs (M. S. Madjarska et al. 2004; S. Yang et al. 2011). This makes the CH boundaries a compelling site for investigating the spatial distribution of MFEs.

This study presents a comprehensive analysis of MFEs using high-cadence (average $\sim$82 s) H$\alpha$ far blue-wing observations from the Big Bear Solar Observatory (BBSO). Since MFEs occur randomly in a large dataset, we adopt an automated detection algorithm, as described in Section 3. Corresponding H$\alpha$ line-center data from the Global Oscillation Network Group (GONG) are used to validate the detections. Statistical analyses of eruption lifetime and size distributions, as well as visualizations of representative events, are presented in Section 4. CH detection is performed using EUV imaging data, and a spatial cross-comparison is conducted between CH regions and MFE locations. Section 5 discusses how these findings can provide new insight into the physical environments that support MFEs and contribute to understanding their connection to small-scale solar activity and heliospheric structures, such as small-scale magnetic flux ropes and switchbacks and Section 6 summarizes the results of this study.

## 2. Data

### *2.1. H$\alpha$ Full-disk Observation*

The observations of MFEs are carried out using full-disk H$\alpha$ blue-wing (6562.3 Å) data from the BBSO H$\alpha$ (FDHA)

instrument (C. Denker et al. 1999). These observations were taken on 2020 June 9 between 15:41:22 UT and 19:57:13 UT, containing full-disk solar images in the H$\alpha$ blue wing (centered at 6562.8–0.8 Å offset). The data was recorded with the JAI TM-4200GE CCD camera, which offers a 2048 × 2048 pixels field of view with a spatial resolution of $\sim 1''$ pixel$^{-1}$. Note that 188 frames were captured during that time frame in patrol mode with an average cadence of ∼83 s and an exposure time of 22 ms.

To confirm the presence of solar filaments at eruption sites, this study uses the H$\alpha$ line-center images (6562.8 Å) from the National Solar Observatory (NSO/GONG) H$\alpha$ Network (F. Hill et al. 1994) in a total of 239 frames on 2020 June 9 from 15:41:50 UT to 19:57:50 UT with the 2048 × 2048 pixels field of view.

### 2.2. Coronal Hole Data

CH boundaries were identified using the Coronal Hole Identification via Multithermal Emission Recognition Algorithm (CHIMERA), an automated and validated method for segmenting coronal holes from SDO/AIA EUV imagery (M. Reiss et al. 2014; T. M. Garton et al. 2018). CHIMERA processes multithermal AIA data across the 171, 193, and 211 Å channels, analyzing intensity ratios to infer relative temperature and density—key discriminants of coronal holes (T. M. Garton et al. 2018).

The analysis uses full-disk Atmospheric Imaging Assembly (AIA) 171/193/211 images from 2020 June 9 around 18:00:00 UT in 4 hr, matched with cotemporal HMI line-of-sight magnetograms for polarity validation. CHIMERA segments candidate dark pixels that exhibit a low temperature, low density, and unipolar magnetic field, thereby excluding filament channels or quiet-Sun depressions that may appear dark in a single wavelength (T. M. Garton et al. 2018). Morphological filtering and a minimum area constraint (typically ⩾1000 arcsec$^2$) are applied to remove spurious noise and small non-CH features. The result is a robust map of coronal holes for that observation time. In total, eight CH regions were identified. CHIMERA's validation against manually curated maps and its stability over time support its use in linking MFE occurrences with open-field regions and CH boundaries.

## 3. Methods

### 3.1. Feature Detection

This study utilizes a custom implementation of the Yet Another Feature Tracking Algorithm (YAFTA), a software package whose core downhill labeling algorithm was originally developed by B. T. Welsch & D. W. Longcope (2003) and which was formally detailed as a tracking suite by C. E. DeForest et al. (2007), to identify MFEs based on the H$\alpha$ blue-wing data. YAFTA is a region-growing algorithm based on thresholding and spatial connectivity and widely used for detecting magnetic and dynamic features on the solar disk (A. Veronig et al. 2000; B. T. Welsch & D. W. Longcope 2003). This algorithm has been adapted for intensity thresholding in the chromospheric H$\alpha$ blue wing, allowing for the identification of compact, dark erupting features (L. D. Krista & P. T. Gallagher 2009). This study adapted YAFTA for intensity-based feature segmentation in chromospheric images.

The detection algorithm follows these key steps: (1) Image smoothing is applied using a Gaussian kernel to suppress noise (see Figures 1(a) and (e)). (2) A static threshold based on background contrast is used to isolate dark features (see Figures 1(b) and (f)). (3) Size filters are applied to exclude unresolved detections (e.g., <10.15 arcsec$^2$ in size). (4) Matching process is applied based on the temporal and spatial position of detected features to identify the same features in different frames (M. Qu et al. 2005). (5) An error-detecting algorithm is applied to detect dark features based on the feature growth rate, size change, and feature movement over time. (6) For each detected eruptive event, the corresponding temporal and spatial position in GONG H$\alpha$ line-center data is checked to ensure the presence of minifilaments. The final catalog contains only those events that passed both detection and verification criteria.

Each image frame $I(x, y)$, where $x$ and $y$ denote coordinates in arcseconds, is processed to extract binary masks corresponding to dark features. A threshold $T$ based on the H$\alpha$ blue-wing images is applied to generate a binary mask $M(x, y)$, where pixels below this threshold are considered part of a feature (see Figures 1(c) and (g)). After thresholding, a connected-component labeling algorithm identifies spatially contiguous regions, each treated as a candidate feature (see Figures 1(d) and (h)). Each feature $F_i$ is assigned basic properties such as its label $L_i$, time step $t_i$, pixel count $N_i$, centroid location $(x_i, y_i)$, and bounding box $B_i$. These properties are used to track the temporal evolution of features across frames. The algorithm is then employed to extract these properties and follow the features over time. The process of binary masking and labeling, including thresholding and feature identification, is illustrated in Figure 1 with the threshold $= \mu - \sigma = -285.3$.

For each labeled region of the mask $M(x, y)$, its physical characteristics were extracted using region-based measurements. To remove noise, features with area $A_i < A_{\min}$ (e.g., 10.15 arcsec$^2$) were excluded. Additionally, to ensure relevance to the activity of the disk, features outside the solar disk (defined by the center $(x_0, y_0) = (1025\overset{''}{.}536, 1025\overset{''}{.}536)$, $R = 946\overset{''}{.}710$) were excluded based on radial distance. However, this algorithm only considered three-fourths of the solar disk to neglect the limb parts of the Sun. This strategy avoids the effects of limb darkening and increasing noises near the limb.

To track MFEs across frames, features in consecutive time steps were matched using centroid proximity. Let $F_i^{(t)}$ and $F_j^{(t+1)}$ detonate features in frames $t$ and $t+1$, respectively. A match will be assigned if the distance between $F_i^{(t)}$ and $F_j^{(t+1)}$ is less than $d_{\max}$, which is the maximum displacement threshold. Matched features inherit the same label, and unmatched features receive new unique labels. This tracking allows us to compute feature lifetimes by counting the number of consecutive frames in which a label persists.

### 3.2. Eruption Analysis

To determine whether the detected feature represents the MFE or not, this study uses a custom postprocessing routine defined by two steps: (1) feature group analysis and (2) eruption certainty scoring, applied to the detected features.

Tracked features with shared labels across multiple time frames were grouped into eruption candidates. Each candidate eruption is defined as a set of features $\{F_k^{(t_1)}, F_k^{(t_2)}, \ldots, F_k^{(t_n)}\}$ associated with the same label and ordered in time. From each group, there is duration or lifetime $\Delta t_k = t_n - t_1$, growth rate $G_k = \frac{A^{(t_n)} - A^{(t_1)}}{\Delta t_k}$, and net displacement or movement $D_k = \sqrt{(x_n - x_1)^2 + (y_n - y_1)^2}$. For a feature group to be considered an eruption candidate, each feature group must satisfy the following conditions. (1) Minimum lifetime: only features that persist for at least 7 minutes are considered. (2) Net growth or motion: the feature must show measurable spatial evolution either $|A^{(t_n)} - A^{(t_1)}| > A_{\text{thresh}}$ or $D_k > D_{\text{thresh}}$. We established $A_{\text{thresh}} = 10$ pixels ($\approx 10\,\text{arcsec}^2$) and $D_{\text{thresh}} = 5$ pixels ($\approx 5''$). Rather than being arbitrarily chosen, these thresholds were empirically determined based on the observational limits of our dataset and the spatial resolution of the tracking algorithm. They are set as small as possible to maximize our sensitivity to the rapid growth or significant macroscopic movement characteristic of MFEs, while remaining strictly above the pixel-level noise floor caused by seeing-induced jitter and instrument fluctuations. These provide basic kinematic descriptors of each eruption candidate.

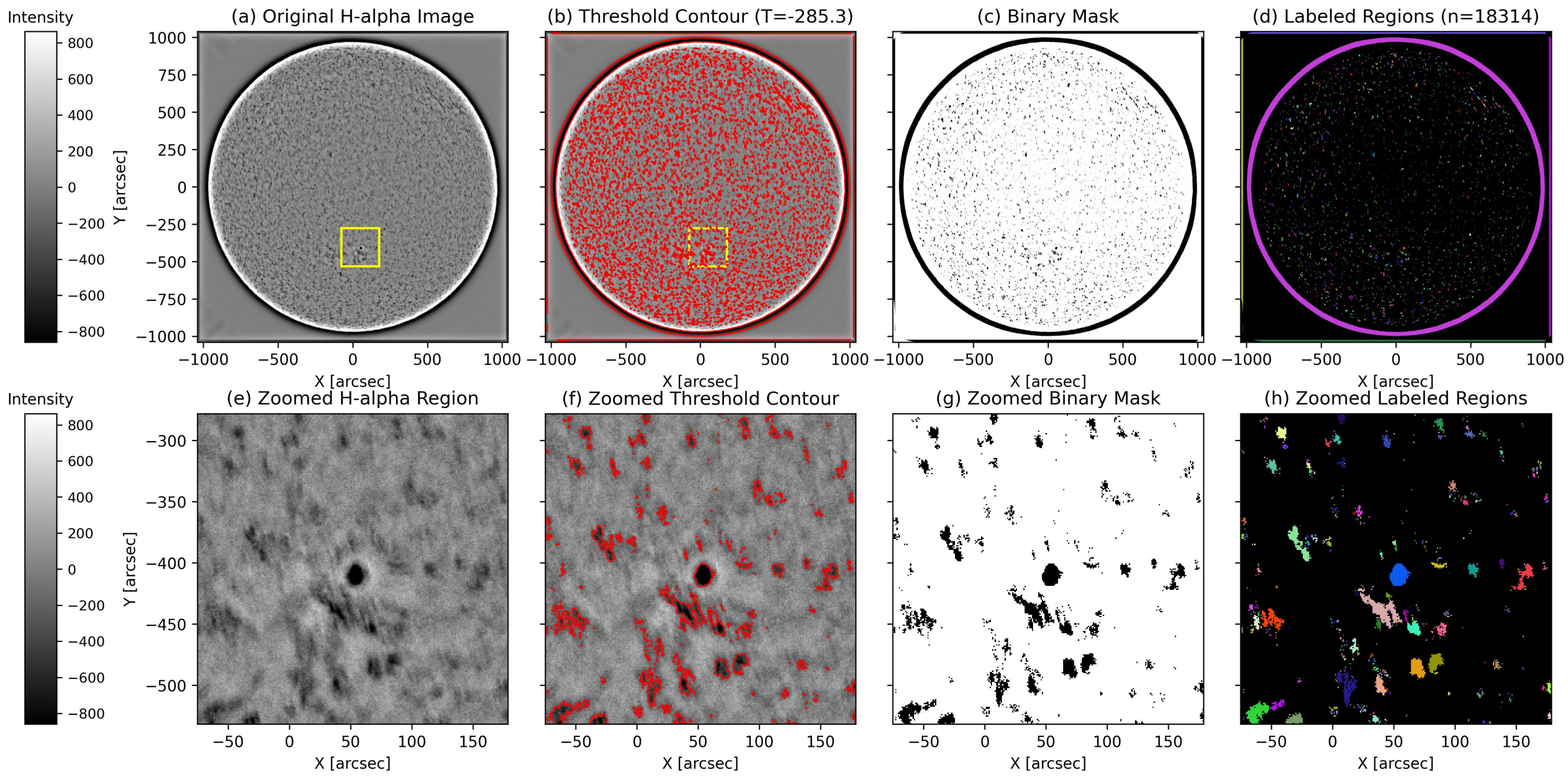

**Figure 1.** The binary masking process for feature detection. (a) Original H$\alpha$ filtergram showing dynamic solar features, captured on 2020 June 9 at 17:51:05 UT. The yellow rectangle indicates the region magnified in the bottom row. (b) Visualization of the threshold contour (red line) applied to identify features. (c) Binary mask created by applying the threshold value of $\mu - \sigma$, where $\mu$ is the mean intensity and $\sigma$ is the standard deviation. (d) Labeled mask showing distinct contiguous regions identified as individual features for tracking across successive frames. Different colors are randomly assigned merely to visually distinguish features with different labels from one another. (e)–(h) The same processing sequence as in panels (a)–(d) in the selected region in panel (a), providing detail of individual features.

To prioritize the most plausible filament eruptions, each candidate was assigned a certainty score $S_k \in [0, 1]$, calculated based on the MFE size. Let $A_{\max}^{(k)}$ be the largest area across all frames in eruption $E_k$, then

$$S_k = \min\left(1.0, \frac{\log_{10}(A_{\max}^{(k)})}{\log_{10}(10^5)}\right). \quad (1)$$

A logarithmic scaling is used here to handle the wide range of sizes, and this gives a normalized confidence score between 0 and 1. For further analysis, the length of each minifilament was defined as the diameter of its bounding box $B$, where $B$ spans from $(x_{\min}, y_{\min})$ to $(x_{\max}, y_{\max})$. The filament length was, thus, computed as the diagonal distance across this box, representing the largest linear extent of the feature. This definition provides a consistent and geometry-independent measure of eruption length across all events, offering a more reliable proxy for filament extent than the projected area alone.

### 3.3. Evaluation

To evaluate the reliability of detected MFEs, this study performed a cross-comparison with cotemporal H$\alpha$ line-center observations from GONG, which can reveal minifilament structures supporting upward motions seen in the blue wing (A. Ahmadzadeh et al. 2024). A two-step preprocessing was applied to account for differences in spatial scale and cadence between datasets: GONG images were rescaled to match the solar disk radius of BBSO images and were temporally aligned with the closest BBSO frame. Validation was performed by checking for dark line-center features within a $10''$ radius around each blue-wing detection's centroid. An example is shown in Figure 2, where the presence of a cospatial dark feature in the line center confirms the eruptive nature of the event.

## 4. Results

The detection and filtering pipeline initially extracted numerous dynamic features from the H$\alpha$ blue-wing dataset. Among these, candidates exhibiting coherent outward motion and temporal expansion over multiple frames were flagged for further analysis. To investigate the typical morphology and dynamics of these events, four representative eruptions were selected. These events were chosen based on a combination of high certainty scores and spatial clarity, which was assessed by considering their well-defined structure, minimal overlap with other features, and visibility throughout the event duration.

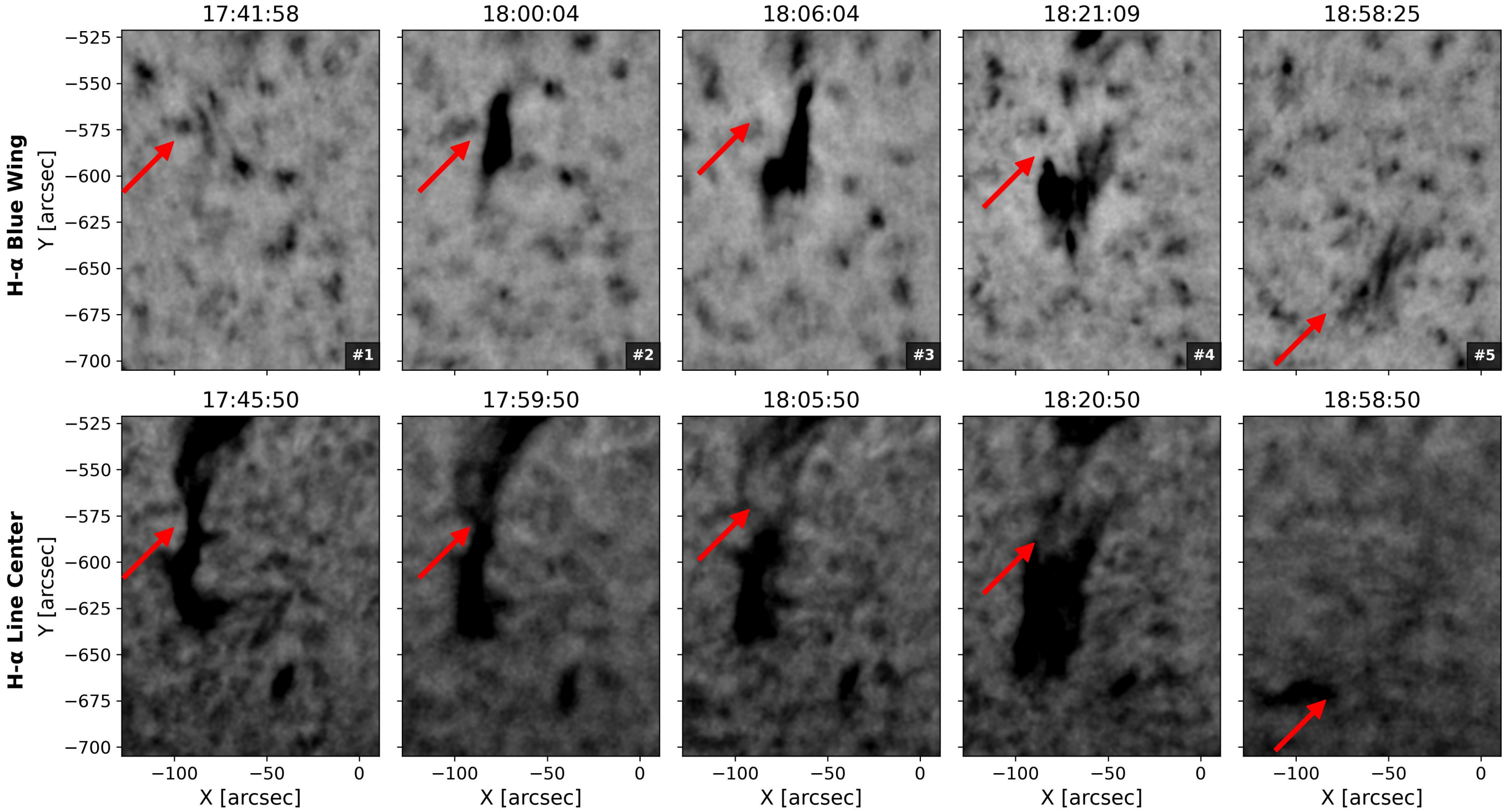

**Figure 2.** Example of the H$\alpha$ line-center validation process for a H$\alpha$ blue-wing-detected eruption. The top row shows H$\alpha$ blue-wing images (taken at a wavelength of H$\alpha$ − 0.8 Å) with the detected MFE and its evolution through time. The bottom row shows the corresponding H$\alpha$ line-center image (taken at a wavelength of 6562.8 Å) at the same time. A nearby darkened feature within the matching radius confirms the eruption in both datasets.

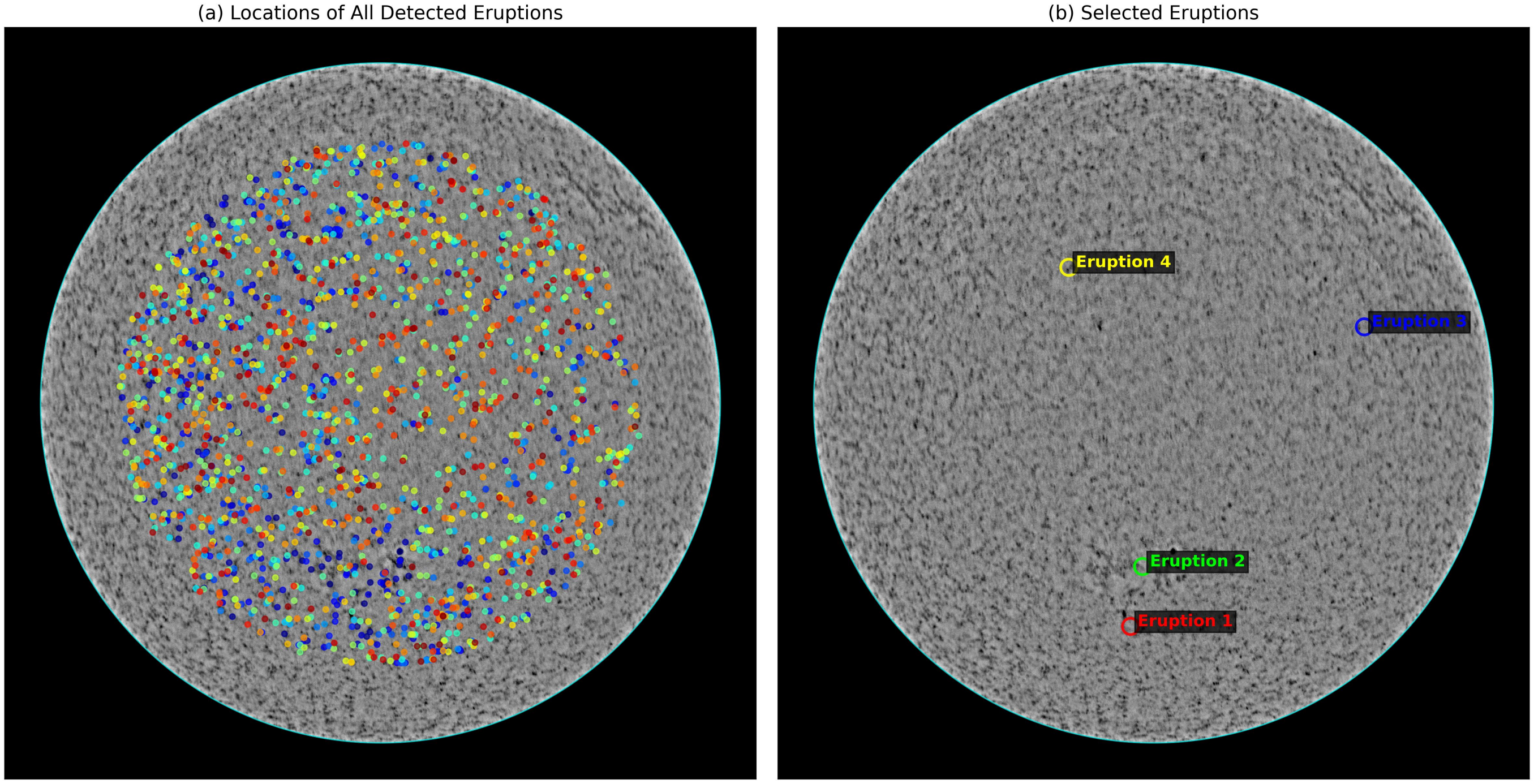

**Figure 3.** Locations of the eruptions marked on an H$\alpha$ image. (a) Spatial location of all detected eruptions within the processed area for the entire dataset. (b) Spatial location of four representative eruptions on the full solar disk (see Figures 4, 5, 6, and 7).

The selection also aimed to capture diversity in eruption properties, including variations in size, lifetime, and location—spanning both AR and quiet-Sun environments. Figure 3 shows the spatial locations of all detected eruptions (Figure 3(a)) and the four selected eruptions (Figure 3(b)), overlaid on a full-disk H$\alpha$ blue-wing frame. In Figure 3(a), each of the detected eruptions is located by a dot on the full-disk image. Each of the four eruptions in Figure 3(b) is labeled with a number that represents the location of detection of the eruptions whose analysis is available in Figures 4, 5, 6, and 7,

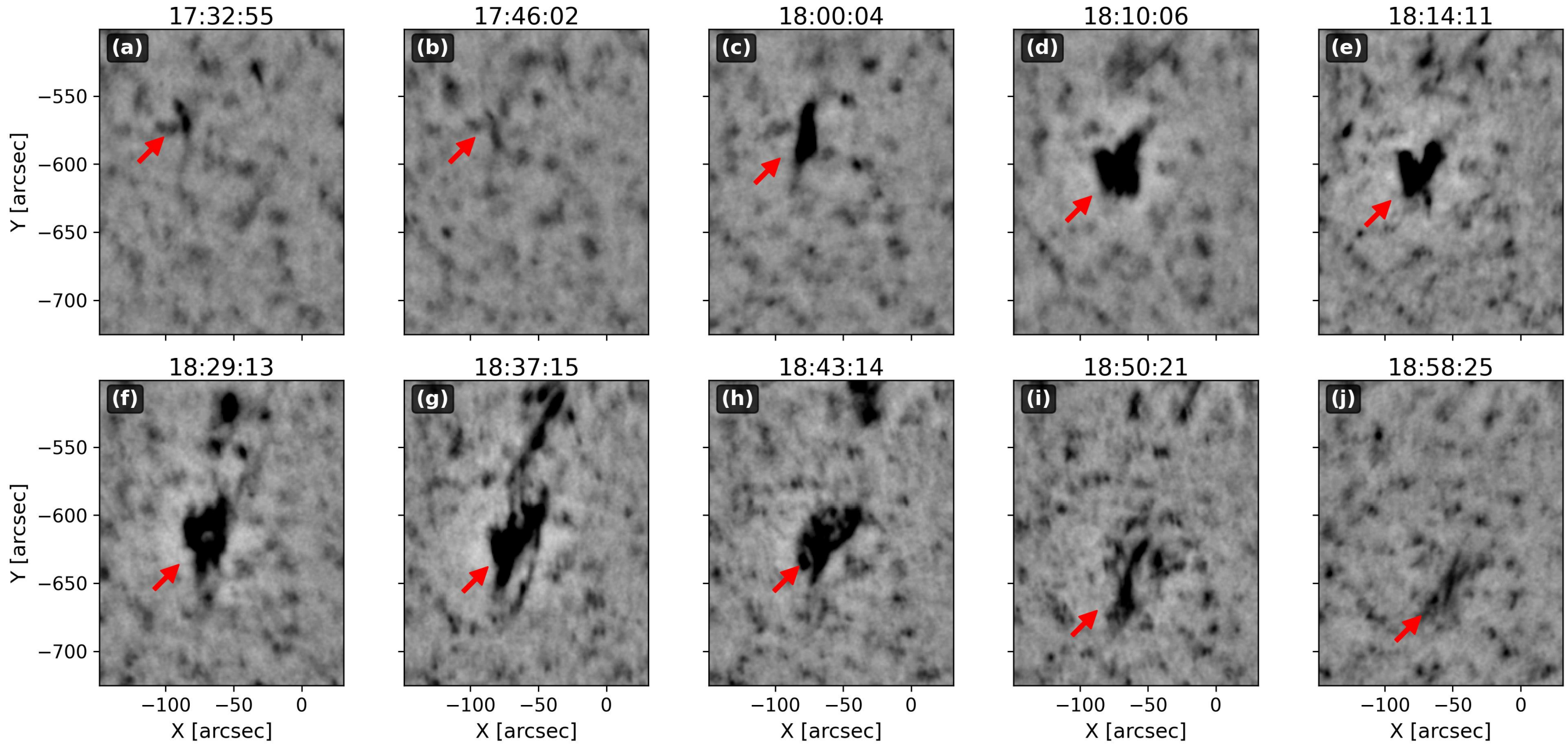

**Figure 4.** Temporal evolution of the largest detected MFE in the Hα blue-wing dataset. The event lasted approximately 137.7 minutes, beginning at 17:28:52 UT and ending at 19:46:34 UT. Red arrows point to the erupting feature in each of the 10 selected frames. The eruption underwent substantial growth, reaching a length of 97.16 Mm, with visible structural elongation and consistent darkening indicative of rising chromospheric material. (The animated version of this figure presents the continuous temporal evolution of the MFE over its full observation period. In the animation, the data is displayed in a side-by-side format: the left panel shows the Hα blue-wing sequence, while the right panel shows the cotemporal Hα line-center sequence. The exact observation time is updated dynamically at the top of the panels. To trace the eruption, a red contour outlines the detected MFE in the blue-wing panel; this identical contour is simultaneously plotted at the exact same spatial coordinates in the line-center panel to validate the corresponding eruptive feature across both datasets).

(An animation of this figure is available in the online article.)

respectively. The selected eruptions occurred at various positions across the solar disk and at different times throughout the observation period, suggesting no spatial or temporal bias in the detection algorithm (see Figure 3).

### *4.1. Eruption Evolution Case Studies*

To qualitatively analyze these MFEs' evolution, 10 frames were selected from each event sequence. These frames capture the main stages of the eruption: initial emergence, growth, peak structure, decay, and dissipation or motion. In each sequence, the erupting feature is pointed with a red arrow.

Figure 4 presents the evolution of a normal-sized filament eruption, characterized by a long, filament-like structure that expands laterally over time. The total area of the absorption enhancement increases significantly. This event spans a total of 103 frames, corresponding to a duration of approximately 137.7 minutes, and covers a length of 97.16 Mm. The eruption initiates as a compact dark structure and rapidly expands both laterally and vertically across the 10 representative frames. The dynamics indicate gradual displacement and morphological coherence throughout this event's evolution, a characteristic often associated with organized magnetic structures. Such coherent motion is consistent with the behavior of flux-rope eruptions or confined filament lift-off, as previously described in theoretical and observational studies of filament dynamics (N. E. Raouafi 2009; S. Yang et al. 2014).

Figure 5 presents another AR filament eruption tracked over 44 frames between 18:56:22 UT and 19:57:13 UT. The event begins as a small dark patch near a sunspot and shows progressive growth in size and elongation, reaching a length of 40.00 Mm. The elongation appears to be oriented toward the northeast, rather than expanding symmetrically. This directional evolution may reflect the influence of the surrounding magnetic topology or anisotropic plasma conditions. The feature fades gradually in the final frames, consistent with the dissipation or cooling of the erupting material.

A more transient eruption is illustrated in Figure 6, representing one of the shortest-duration events detected in the dataset. This eruption lasted only 9.1 minutes, beginning at 16:09:28 UT and ending at 16:18:32 UT, with a total of nine tracked frames. Despite its brief lifetime, the feature reached a maximum size of 27.94 Mm before rapidly fading. The final frame in the sequence shows the complete disappearance of the filament, indicating that the structure either dissipated or erupted fully out of the chromospheric layer. This example demonstrates the algorithm's sensitivity to short, impulsive events that may otherwise go undetected in coarser datasets.

An MFE located in the northeastern quadrant of the solar disk is shown in Figure 7. This event persisted for 53 frames, spanning from 16:14:29 UT to 17:22:56 UT, and displayed consistent growth throughout its evolution. The feature originated near coordinates (791.″3, 1393.″2) and expanded to 26.77 Mm over its lifetime. This eruption demonstrated steady expansion. Its evolution showcases typical characteristics of an MFE—including expansion, motion, and eventual fading—making it an ideal representative of structured, mid-duration eruptions with moderate spatial scale.

These case studies demonstrate that the detection algorithm can reliably capture a variety of eruption morphologies, durations, and dynamical behaviors, ranging from compact, transient absorption enhancements to long, evolving filamentary structures.

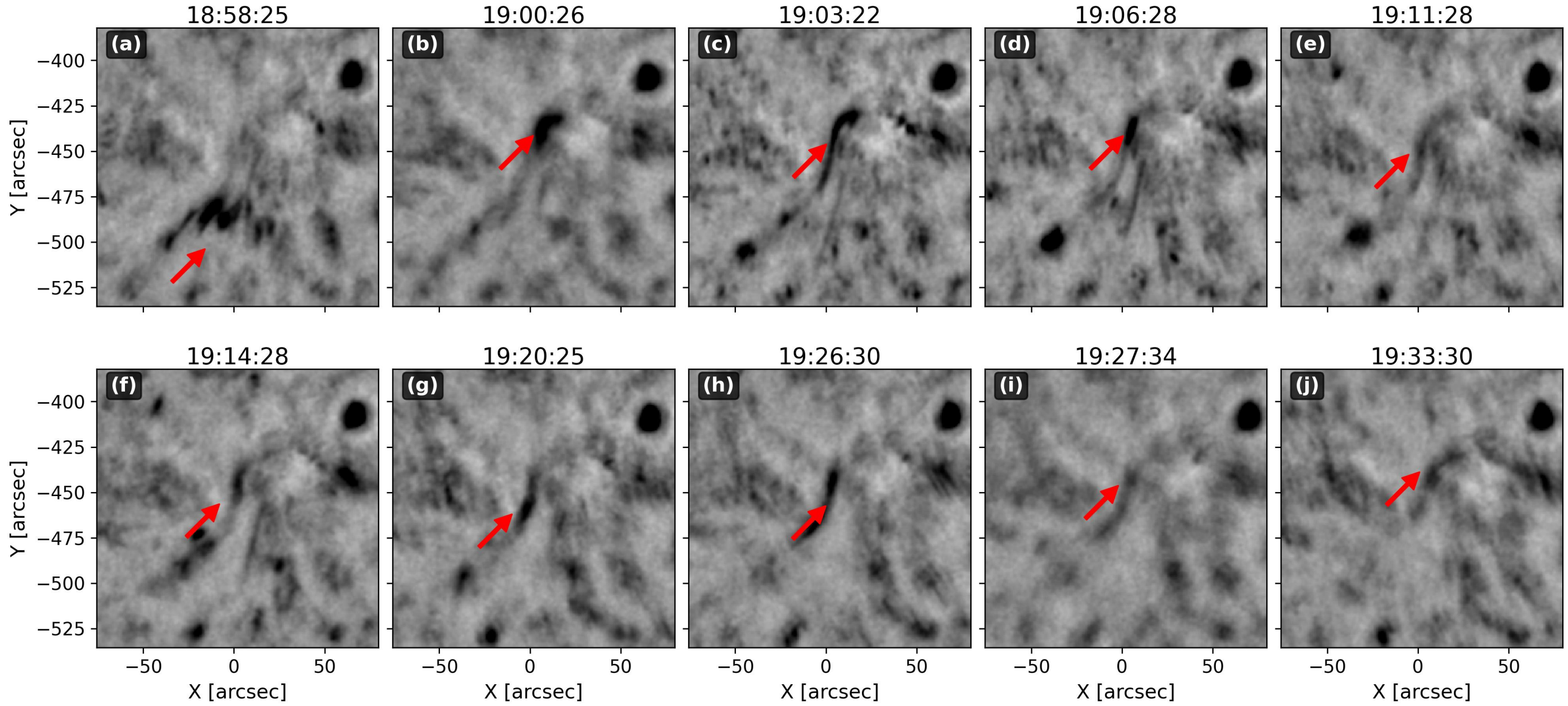

**Figure 5.** Evolution of a representative MFE observed in the H$\alpha$ blue-wing images. This event occurred between 18:56:22 UT and 19:57:13 UT, spanning approximately 60.9 minutes. The red arrows indicate the location of the erupting feature in each of the 10 selected frames. The eruption starts as a compact darkening, gradually elongates, and then fades as the material dissipates or lifts off. The filament shows signs of directional motion and partial dissipation over time, suggesting a structured eruptive behavior. (The animated version of this figure presents the continuous temporal evolution of the MFE over its full observation period. In the animation, the data is displayed in a side-by-side format: the left panel shows the H$\alpha$ blue-wing sequence, while the right panel shows the cotemporal H$\alpha$ line-center sequence. The exact observation time is updated dynamically at the top of the panels. To trace the eruption, a red contour outlines the detected MFE in the blue-wing panel; this identical contour is simultaneously plotted at the exact same spatial coordinates in the line-center panel to validate the corresponding eruptive feature across both datasets).

(An animation of this figure is available in the online article.)

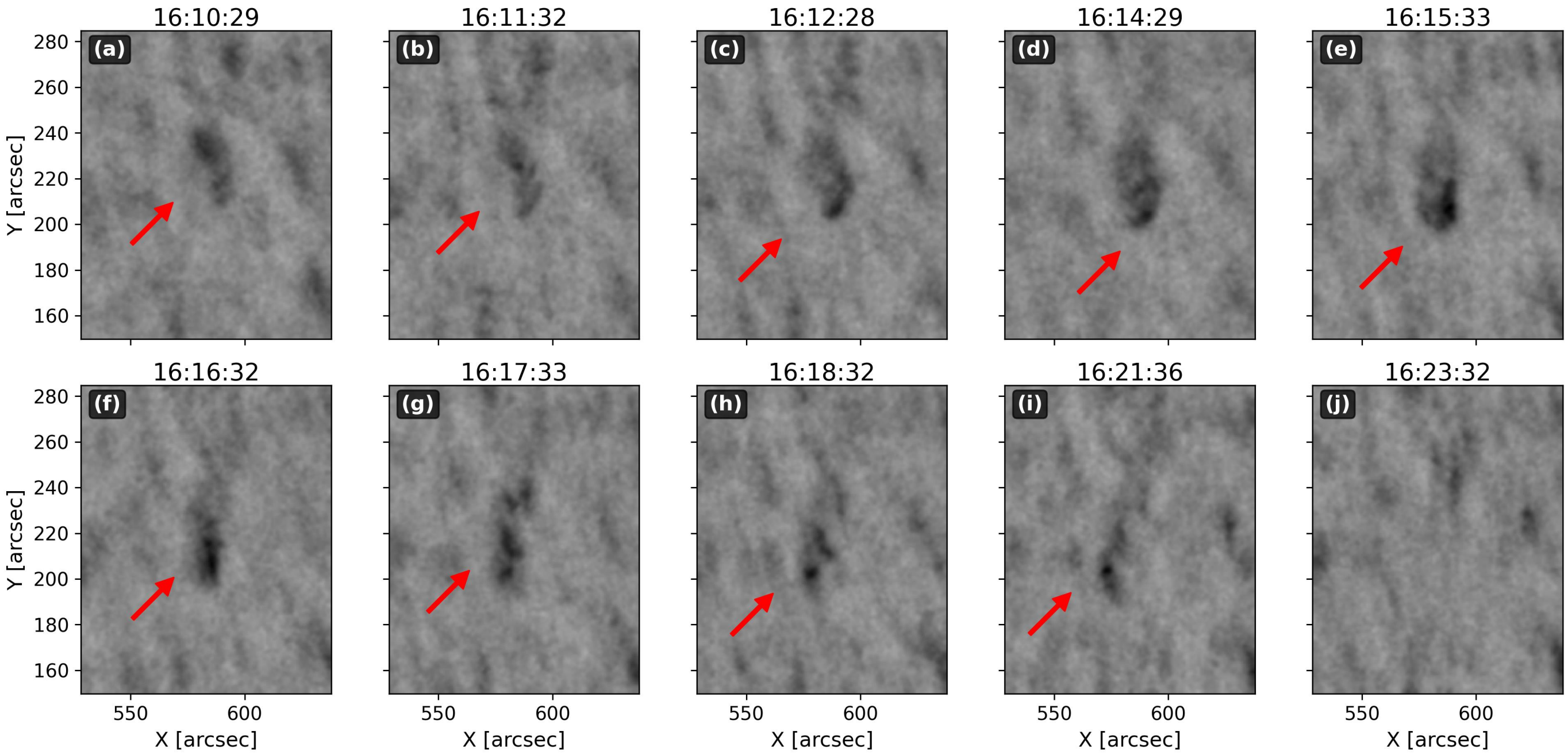

**Figure 6.** Temporal evolution of a short-lived MFE observed in the H$\alpha$ blue wing. The eruption spans from 16:09:28 UT to 16:18:32 UT, with a total duration of 9.1 minutes and a maximum size of 27.94 Mm. The red arrows mark the location of the feature across 10 selected frames. The filament exhibits rapid growth followed by a sudden disappearance, as can be seen in the final panel, indicating a complete vanishing of the structure. This event highlights the detection method's ability to capture small-scale, impulsive chromospheric dynamics. (The animated version of this figure presents the continuous temporal evolution of the MFE over its full observation period. In the animation, the data is displayed in a side-by-side format: the left panel shows the H$\alpha$ blue-wing sequence, while the right panel shows the cotemporal H$\alpha$ line-center sequence. The exact observation time is updated dynamically at the top of the panels. To trace the eruption, a red contour outlines the detected MFE in the blue-wing panel; this identical contour is simultaneously plotted at the exact same spatial coordinates in the line-center panel to validate the corresponding eruptive feature across both datasets).

(An animation of this figure is available in the online article.)

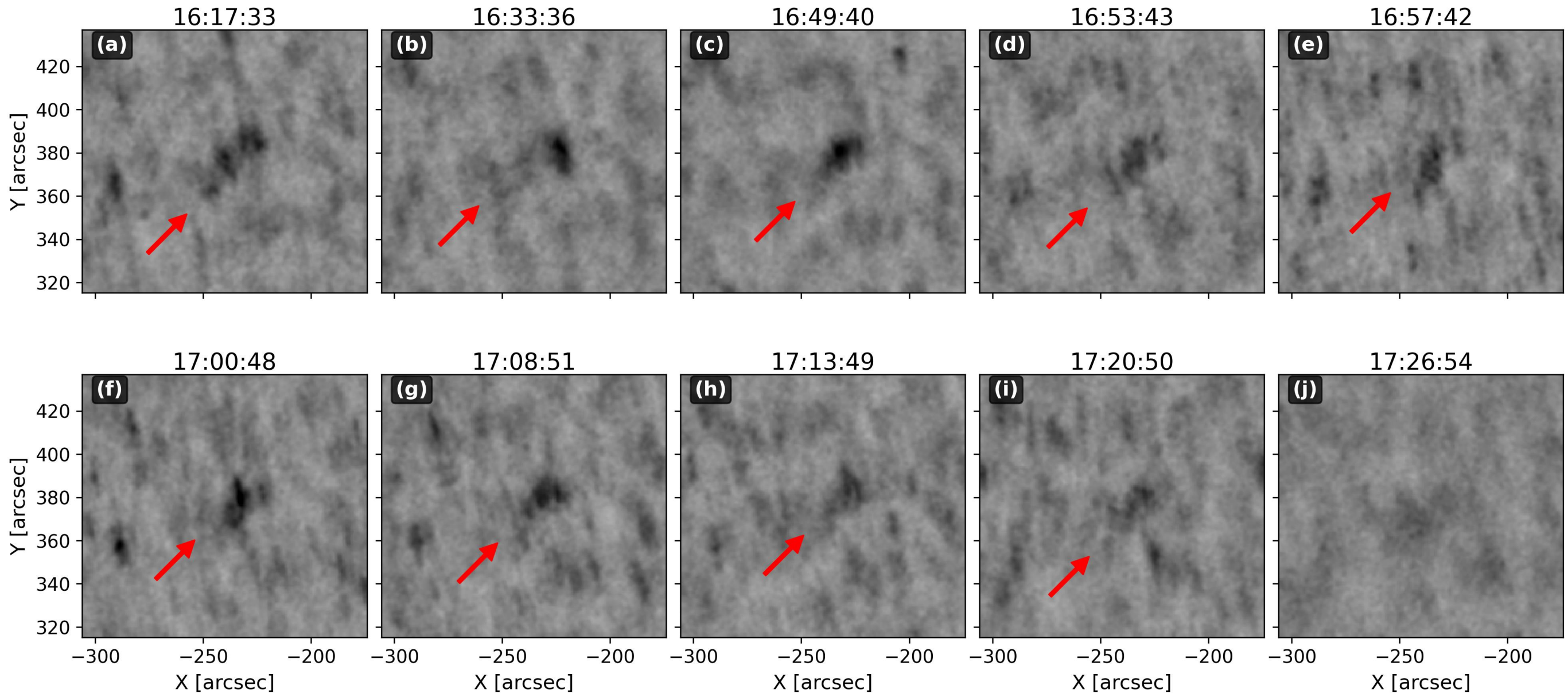

**Figure 7.** Temporal evolution of an MFE in the northeastern quadrant of the solar disk, tracked from 16:14:29 UT to 17:22:56 UT, spanning about 68 minutes. Red arrows highlight the erupting structure in each selected frame. The feature originated near coordinates (791″.3, 1393″.2) and expanded to 26.77 Mm. (The animated version of this figure presents the continuous temporal evolution of the MFE over its full observation period. In the animation, the data is displayed in a side-by-side format: the left panel shows the Hα blue-wing sequence, while the right panel shows the cotemporal Hα line-center sequence. The exact observation time is updated dynamically at the top of the panels. To trace the eruption, a red contour outlines the detected MFE in the blue-wing panel; this identical contour is simultaneously plotted at the exact same spatial coordinates in the line-center panel to validate the corresponding eruptive feature across both datasets).

(An animation of this figure is available in the online article.)

### *4.2. Statistical Distribution of MFE Properties*

After analyzing the lifetime and size distribution of MFEs, we implemented another strategy in this research. To ensure a robust statistical analysis of the power-law tail, lower and upper thresholds were applied to the dataset. The lower size threshold was set at an area of 67 arcsec$^2$, which corresponds to the peak of the observed area distribution. In astrophysical power-law distributions, this peak defines the statistical completeness limit (or turnover point) of the observations. The positive slope below 67 arcsec$^2$ does not imply a lack of smaller MFEs, but rather reflects the drop in detection efficiency due to the spatial resolution limits of the data and the algorithmic tracking constraints. By restricting the slope analysis to events larger than 67 arcsec$^2$, we ensure the power-law fits are applied only to the fully resolved, statistically complete portion of the MFE population. Furthermore, upper limits on size and lifetime were defined empirically by the largest single event observed in our dataset (the 97 Mm, 137 minutes eruption detailed in Section 4.1). Because structures of this magnitude transition into the regime of standard, large-scale filaments, this event serves as a natural, data-driven upper boundary for isolating the minifilament parameter space. After applying size-based eruption criteria and cross-validating with Hα line-center observations, a total of 1986 MFE candidates were cataloged. This number will lead us to calculate the occurrence rate of MFEs in ∼4 hr of observation:

$$\begin{aligned}\text{Occurrence Rate} &= \frac{\text{Number of detected MFEs}}{\text{Processing Area} \cdot \text{Time Interval}} \\ &= \frac{1986}{(1.06 \cdot 10^{6}\ \text{Mm}^2) \cdot (4.26\ \text{hr})} \\ &\approx 4.41 \times 10^{-4}\ (\text{Mm}^2 \cdot \text{hr})^{-1}. \qquad (2)\end{aligned}$$

Assuming a uniform distribution across the solar surface, this translates to a global occurrence rate of $\sim 6.61 \times 10^4$ MFEs per day and $\sim 1.12 \times 10^4$ at the processed area.

This study calculates the lifetime and size distribution of MFEs to better characterize their occurrence patterns and investigate potential scaling relationships in their spatial and temporal evolution. The statistical distribution of their lifetimes and lengths, as well as the relationship between these two parameters is shown in Figure 8 (see, for better visualization, the large-scale filament mentioned in Figure 4). The lifetime distribution (Figure 8(a)) shows a steep decline in frequency with increasing duration. Most eruptions lasted <30 minutes, with a median duration of 12.1 minutes and a mean of 20.6 ± 23.0 minutes. A small number of events, however, persisted beyond 100 minutes, forming a long-tailed distribution indicative of rare, long-lived filaments or minifilaments. Some of these long-lived detections may be attributed to light broadening rather than true eruptive activity; however, such cases are rare and constitute a negligible fraction of the dataset, having no significant impact on the statistical trends observed in this study. The length distribution (Figure 8(b)) also follows a steep decay. The median minifilament length is 15.5 Mm, while the mean is 16.6 ± 4.3 Mm.

The histogram is shown in log–log space, revealing a steep decay. To address the statistical biases inherent in least-squares fitting of power-law distributions on binned data (M. L. Goldstein et al. 2004; H. Bauke 2007), we evaluated the scaling behavior using maximum likelihood estimation (MLE) on the unbinned data. This declining regime is measured starting from the peak of the distribution (the completeness limit) at 14.28 Mm. To ensure the fit strictly characterizes the core minifilament population, we applied an upper physical boundary of 36.0 Mm to the MLE calculation. This cutoff prevents extreme outliers—which physically transition into the regime of standard, large-scale filaments—from artificially skewing the statistical lever arm. The resulting main MLE-derived slope for this MFE population is −4.78. It should be noted that at the extreme tail end of the plotted distribution (>36 Mm), the histogram bins visually deviate from the fitted line. This variance is a standard artifact of Poisson noise (small number statistics). Because these massive eruptions are exceedingly rare, bins in this region contain only a few discrete events. The logarithmic scaling of the *y*-axis severely exaggerates these discrete counts, whereas the unbinned MLE approach correctly models the underlying probability density without being biased by this low-count binning noise. While individual events vary widely, there is a clear trend that larger eruptions tend to last longer. It is worthwhile to establish the minimum spatial scale required for an MFE to survive for a given duration. To quantify this threshold, we applied an empirical lower-envelope fit to the data (Figure 8(c)). The data domain was divided into 10 uniform temporal bins, and the local minimum length within each bin was extracted. An ordinary least-squares (OLS) linear regression was then performed on these boundary points in log–log space, yielding the power-law relation:

$$\text{Length} \propto \text{Duration}^{0.22}. \tag{3}$$

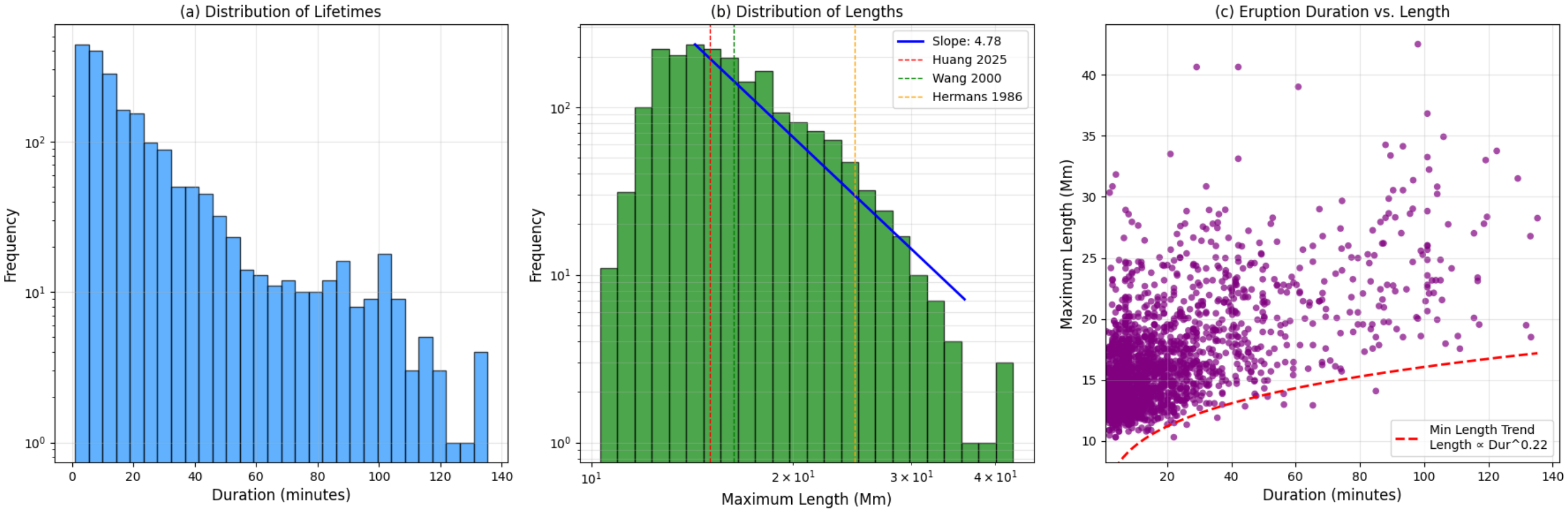

**Figure 8.** Statistical properties of the 1986 MFEs detected in the H$\alpha$ blue-wing dataset. (a) Histogram of eruption durations (in minutes), plotted on a log scale. Most events are short lived, with a steep drop-off beyond 30 minutes. (b) Histogram of maximum eruption lengths (in Mm) on log–log axes. The solid line represents a single, continuous power-law fit derived using unbinned maximum likelihood estimation (MLE). To strictly isolate the minifilament regime, the fit is bounded between the completeness limit (14.28 Mm) and an upper physical cutoff (36.0 Mm). The distribution follows a steep decay with a scaling index of −4.78, indicating that the probability of occurrence drops rapidly at larger spatial scales. (c) Scatter plot of duration vs. length. A power-law fit to the lower envelope (red dashed line) shows a sublinear trend size $\propto$ duration$^{0.22}$, indicating that larger events tend to last longer, but grow more slowly than linearly.

This sublinear scaling suggests that longer-lived eruptions grow in length, but not proportionally with time, possibly due to geometric or magnetic confinement. It is important to note that this fit is applied to the lower envelope of the distribution, rather than to the mean trend. In other words, the relation estimates the *minimum length required for a filament to survive for a given duration*, defining a physical "survival boundary" rather than the central scaling of the population. Thus, the inferred scaling should be interpreted as a constraint on the smallest detectable eruptions at each timescale, rather than a universal growth law for all events. It should be noted that the fit visually diverges from the boundary for durations about <10 minutes. This deviation is not a failure of the scaling law, but rather an artifact of observational limits: at durations under 10 minutes, the data becomes heavily quantized by the instrument's ∼83 s cadence, and the lengths reach the spatial detection floor of ∼10 Mm, preventing the observation of a continuous boundary in that regime.

To place these findings in context, Figure 8(b) shows the size distribution obtained in this research, with the detection thresholds corresponding to previous studies overplotted for comparison. Notably, all three of these earlier investigations—N. Huang & H. Wang (2025), J. Wang et al. (2000), and L. M. Hermans & S. F. Martin (1986)—were also based on H$\alpha$ observations from BBSO, although with differing field-of-view and size thresholds.

N. Huang & H. Wang (2025) reported $\sim 1.1 \times 10^4$ MFEs per day, which is ∼6 times lower than the occurrence rate estimated in this research. This difference is expected because this study employed a continuous automated detection algorithm in full-disk images, whereas N. Huang & H. Wang (2025) manually traced features in high-resolution H$\alpha$ images with a very limited field of view. In their study, the reported size range for MFEs was 4″–13″. In contrast, this research adopted a minimum size threshold of 14″. For direct comparison, if the threshold here were increased to 20.″7, matching that in N. Huang & H. Wang (2025), the estimated occurrence rate would become consistent with their result.

A rate of 6200 MFEs per day was reported by J. Wang et al. (2000) for the quiet Sun, using H$\alpha$ images from BBSO with a reported size range of 20″–40″. In this research, if the detection threshold were set to 22.″5, Equation (2) would yield the same occurrence rate as in J. Wang et al. (2000).

On the other hand, L. M. Hermans & S. F. Martin (1986) reported only 600 MFEs per day, also using BBSO H$\alpha$ images but focusing solely on the quiet Sun. They reported a length range of 5″–54″, with an average of 15″. In this research, if the threshold were set to 34.″1, Equation (2) would produce the same rate as in L. M. Hermans & S. F. Martin (1986).

A. C. Sterling & R. L. Moore (2016) presented a three-point power-law distribution of the estimated number of erupting-filament-like features on the Sun at any given time, plotted as a function of the size of those erupting features. The estimated number of MFEs driving coronal jets and jet bright points is $\sim$4 at a size of $\sim$8 Mm. The postulated microfilament eruptions that would drive spicules are as many as $\sim 10^5$ at a size of $\sim$0.3 Mm. On the other hand, J. Lee et al. (2024) analyzed high-resolution H$\alpha$ observations at BBSO of a network boundary near a CH to derive a spicule occurrence rate as high as 0.55 spicules $Mm^{-2}$ $s^{-1}$. Together, these aggregate numbers yield a global power-law distribution for erupting-filament-like features with a slope of $-2.7$ spanning from 0.3–66 Mm. In this research, however, the size distribution of MFEs evaluated via unbinned MLE reveals a significantly steeper, continuous decline. Specifically, for the isolated MFE regime between 14.28 and 36.0 Mm, the measured slope is $-4.78$. While this slope is much steeper than the $-2.7$ index derived by A. C. Sterling & R. L. Moore (2016), the two results describe fundamentally different statistical regimes and are not contradictory. The $-2.7$ index represents an interclass "macroscaling" envelope connecting the average occurrence rates of three distinct physical phenomena (spicules, coronal jets, and CMEs) across nearly 3 orders of magnitude. In contrast, our $-4.78$ index characterizes the intraclass distribution strictly within the minifilament population. This steep internal decay suggests that while minifilaments as a general class are abundant enough to align with the global $-2.7$ scaling, the physical capability of the local magnetic environment to support and erupt progressively larger coherent structures within this specific regime drops off rapidly.

The physical characteristics of the MFEs in our global sample also align well with recent spectroscopic investigations of individual small-scale eruptions. For instance, Y. Kotani et al. (2023) utilized H$\alpha$ imaging spectroscopy to analyze cold plasma ejections in the quiet Sun. While they focused on energetics—using cloud-model fitting to extract precise masses and kinetic energies for 25 selected events—they found typical spatial lengths of $\sim$4–20 Mm, which is in excellent agreement with our statistically derived mean MFE length of 16.6 Mm. While Y. Kotani et al. (2023) established a unified scaling law between ejection mass and flare energy, our study complements this physical picture by providing a larger sample in a statistical context. We demonstrate that such events are not only ubiquitous (occurring globally at a rate of $\sim 6.6 \times 10^4$ per day) but also exhibit distinct morphological confinement laws (our sublinear length–duration scaling) that are heavily modulated by their specific large-scale magnetic environment, such as proximity to coronal holes or ARs.

### 4.3. Length Distribution of MFEs with Respect to an Active Region

To investigate whether MFEs near ARs differ in their physical extent compared to those farther away, this study analyzed the size distribution of eruptions as a function of distance from an identified AR located at $(x, y) = (1057''.3, 619''.6)$ on the solar disk. A radial range of 300″ around the AR centroid was considered, and 312 eruptions were detected within this proximity.

As shown in Table 1, the innermost bin (0″–75″) exhibits the largest average length of 22.4 Mm, with a median of 19.6 Mm, suggesting that larger eruptions tend to occur closest to the AR core. In contrast, the outer bins display smaller average lengths, 17.8 Mm in the 75″–150″ range, 17.5 Mm in 150″–225″, and 16.0 Mm in 225″–300″.

**Table 1**
Size Distribution of Eruptions with Respect to the Active Region

| Distance Range (arcsec) | Eruption Count | Avg. Length (Mm) | Median Length (Mm) | Length Range (Mm) |
|---|---|---|---|---|
| 0–75 | 18 | 22.4 | 19.6 | 12.3–40.6 |
| 75–150 | 58 | 17.8 | 16.4 | 11.4–32.3 |
| 150–225 | 102 | 17.5 | 15.5 | 11.3–97.2 |
| 225–300 | 124 | 16.0 | 14.9 | 10.3–31.4 |
| Active Region | 302 | 17.2 | 15.7 | 10.3–97.2 |
| Full Disk | 1986 | 16.6 | 15.5 | 10.3–97.2 |

These results indicate that eruption size is moderately enhanced in the immediate vicinity of the AR, potentially reflecting the role of magnetic energy concentration in producing more extended eruptions. At the same time, the presence of large events beyond 150″, including one exceeding length of 97.2 Mm, also implies that MFEs are not strictly confined to AR cores and may arise under suitable conditions elsewhere on the solar disk. Although the average length within the 0″–75″ region is also significantly larger than the global average eruption length of 16.6 Mm (see Section 4.2), small-scale eruptions remain abundant near ARs, suggesting that ARs primarily extend the upper end of the size distribution rather than suppressing the occurrence of the smallest events.

### 4.4. Number Density Distribution of MFEs with Respect to Coronal Holes

CHIMERA-derived CH segmentation at 19:00:00 UT was used as the reference map for this analysis. To ensure temporal consistency, the heliographic coordinates of all detected eruptions were rotated to this reference time using solar differential rotation. A few events were displaced beyond the analyzed disk region after rotation and were excluded from the statistics. The analysis covered the inner 69.75% of the solar disk ($1.06 \times 10^6$ $Mm^2$), partitioned into three regions: (1) CH interiors (90 events; 4.6%), (2) CH boundary zones (477 events; 24.3%), and (3) outside CH regions (1392 events; 71.1%).

The eruption density varied across these regions: 1918.0 MFEs per $10^6$ $Mm^2$ outside CHs, 1754.8 MFEs per $10^6$ $Mm^2$ along CH boundaries, and 1530.1 MFEs per $10^6$ $Mm^2$ inside CHs. Relative to the outside regions, this corresponds to $\sim$20% reduction in eruption density inside CHs and $\sim$9% reduction at their boundaries. Figure 9 summarizes these results, showing eruption locations (panel (a)) and a combined analysis of lifetime and size distributions across CH environments (panel (b)).

A breakdown by lifetime category shows that eruptions inside CHs and at their boundaries are predominantly short-lived. Specifically, 78.9% of eruptions inside CHs and 76.5% at boundaries have lifetimes $\leqslant$20.6 minutes, compared with 63.5% outside CHs. Similarly, small eruptions ($\leqslant$16.6 Mm) constitute 73.3% of events inside CHs and 67.1% at boundaries, compared with 61.7% outside CHs. This indicates that CH regions preferentially host shorter and smaller events, consistent with the influence of open magnetic field topology in limiting eruption growth and confinement.

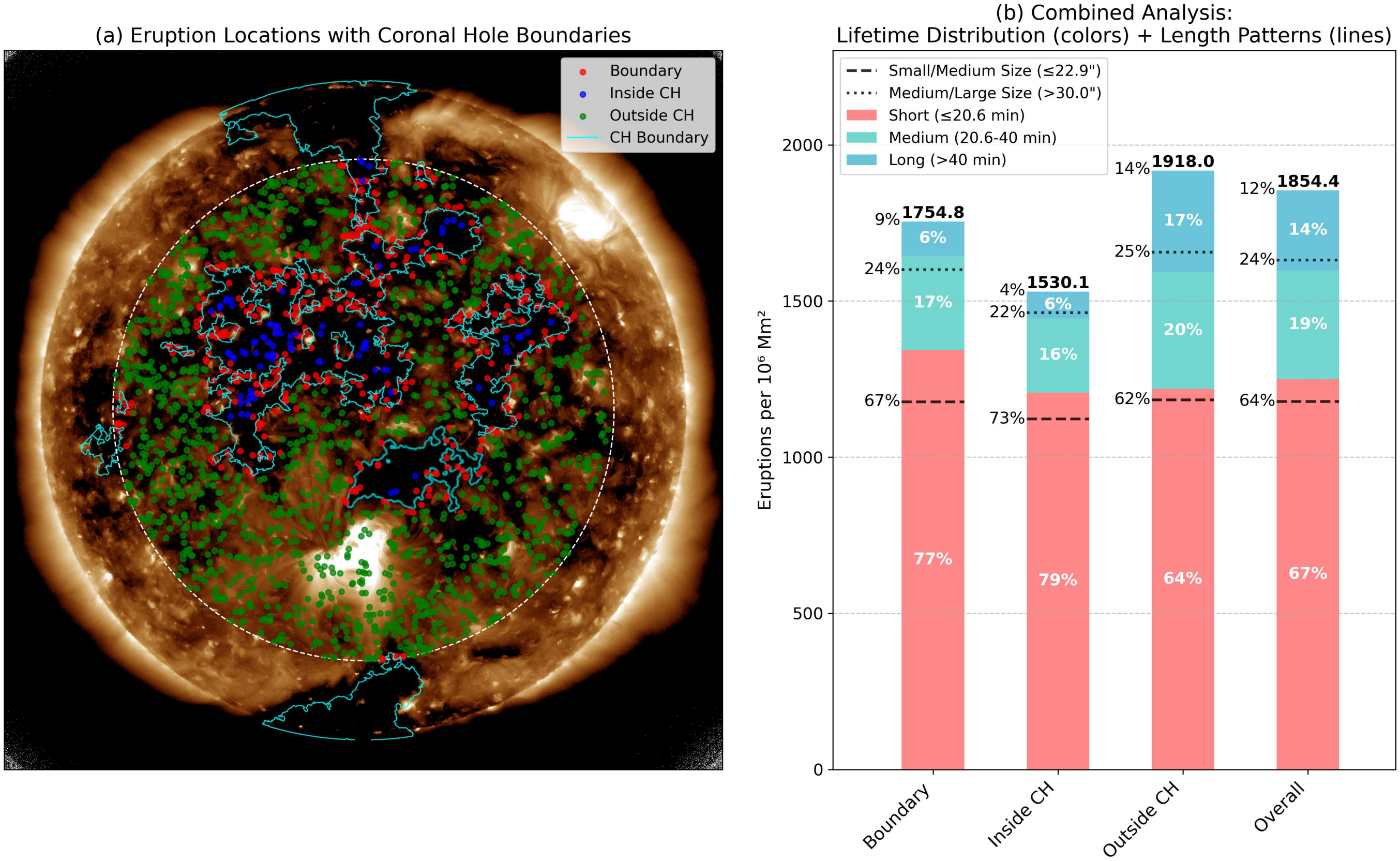

**Figure 9.** Spatial and statistical relationship between minifilament eruptions (MFEs) and coronal hole (CH) boundaries on 2020 June 9. (a) Eruption locations color coded by position relative to CH boundaries, overplotted on an SDO/AIA 193 Å image with CHIMERA-derived CH boundaries (cyan). (b) Combined lifetime (bar colors) and size (dashed/dotted lines) distribution of eruptions in each region. Bars represent lifetime categories: short (⩽20.6 minutes, red), medium (20.6–40.0 minutes, teal), and long (>40 minutes, blue). Dashed and dotted lines mark small/medium and medium/large size thresholds, respectively. Numbers and percentages indicate the proportion of eruptions in each class. The analysis shows that CH interiors and boundaries are dominated by smaller, shorter-lived eruptions, while outside CHs host a greater fraction of large, long-lived events.

To test the statistical robustness of these trends, a series of chi-square and Kruskal–Wallis tests were applied. The Kruskal–Wallis test is a rank-based, nonparametric statistical method used here to verify whether independent samples originate from the same distribution; we use it to confirm that the observed variations in MFE lifetime and length across the three distinct CH regions are statistically meaningful rather than due to random fluctuations (W. H. Kruskal & W. A. Wallis 1952). Both lifetime and size distributions differ significantly across regions (lifetime: $\chi^2 = 46.23$, $p = 2.2 \times 10^{-9}$; length: $\chi^2 = 14.71$, $p = 5.3 \times 10^{-3}$). Nonparametric Kruskal–Wallis tests confirm these differences ($p = 4.7 \times 10^{-11}$ for lifetime and $p = 5.3 \times 10^{-5}$ for length). Following this, pairwise Mann–Whitney tests—a nonparametric method used to compare differences between two independent groups—were applied to determine exactly where these variations occur (H. B. Mann & D. R. Whitney 1947). These tests further show that lifetimes in CHs are significantly shorter than outside ($p = 2.5 \times 10^{-5}$), and lengths are also smaller ($p = 9.7 \times 10^{-5}$). Median lifetimes and lengths were 9.6 minutes and 14.3 Mm inside CHs, compared with 15.1 minutes and 15.6 Mm outside.

These results demonstrate that both the occurrence rate and physical properties of MFEs are measurably affected by the surrounding magnetic environment. CH interiors and boundaries not only show lower eruption density but also exhibit statistically shorter and smaller eruptions, supporting the interpretation that open magnetic fields suppress large-scale development and confine eruption evolution.

## 5. Discussion

Small-scale chromospheric eruptions, including MFEs, represent one of the most frequent and energetically significant modes of magnetic energy release on the solar surface. Recent high-resolution and statistical studies have shown that such eruptions are capable of heating the low corona and loading mass into open magnetic structures through interchange or breakout reconnection. These processes couple the chromosphere and corona on small spatial scales, injecting twist and plasma that can later evolve into fine-scale heliospheric structures.

Observations from N. Huang et al. (2023) demonstrate that numerous small eruptions and blowout jets near CH boundaries can account for a significant fraction of the PSP detections of small magnetic flux ropes (SMFRs). Likewise, the Hα and EUV analyses of N. Huang & H. Wang (2025) confirm that MFEs often act as chromospheric precursors of jet-like eruptions, releasing both cool filament material and hot plasma into the corona. Our global Hα statistics support this picture: this research finds that MFEs are suppressed inside CHs but enhanced along CH boundaries, precisely where open and closed magnetic fields interact. Such boundary eruptions provide natural sites for interchange reconnection that can transfer small flux ropes, twists, or helical field bundles into the nascent solar wind.

Based on the present study, it is worthwhile to estimate the frequency at which the PSP encounters one MFE. Given the strongly skewed lifetime distribution of MFEs, this research adopts the median lifetime ($\tau_{\rm mfe} \approx 12.1$ minutes) as the characteristic source duration. If an erupting MFE injects a coherent magnetic structure into the low corona, its characteristic axial extent can be estimated as (N. Huang et al. 2023)

$$\ell_{\rm mfe} \approx v_{A\odot} \tau_{\rm mfe}, \tag{4}$$

where $v_{A\odot} \approx 1000\,{\rm km\,s^{-1}}$ is the characteristic Alfvén speed near the source region. The corresponding spacecraft-frame passage time of this structure at PSP is then

$$\tau_{\rm cross} \approx \frac{\ell_{\rm mfe}}{v_{\rm sw} + v_{\rm spike}} = \frac{v_{A\odot}}{v_{\rm sw} + v_{\rm spike}} \tau_{\rm mfe} \approx 30 \text{ minutes}, \tag{5}$$

where $v_{\rm sw} \approx 300\,{\rm km\,s^{-1}}$ is the background solar-wind speed, and $v_{\rm spike} \approx 100\,{\rm km\,s^{-1}}$ represents the additional local velocity enhancement associated with the transient. Thus, $\tau_{\rm cross}$ should be interpreted as the expected in situ crossing duration of one MFE-related structure, rather than the lifetime of the MFE on the Sun.

Assuming the one-third aspect ratio, the characteristic MFE width is $w_{\rm mfe} \approx 1/3 \times \tilde{L} = 5.2 \times 10^3$ km. An ejection must exist within a distance less than $w_{\rm mfe}$ to PSP to be detected. The effective cross-sectional area is

$$A'_{\rm PSP\odot} = \pi w_{\rm mfe}^2 + 2 w_{\rm mfe} \cdot v_{\rm PSP\odot} dt, \tag{6}$$

where $v_{\rm PSP\odot}$ is the speed of PSP footpoint moving on the solar surface.

The occurrence rate of MFEs is $S_{\rm mfe} \approx 4.41 \times 10^{-10}\,{\rm km^{-2}\,hr^{-1}}$, which is an order higher than the occurrence rate of small-scale ejections detected in corona, presented in N. Huang et al. (2023).

Using the occurrence rate obtained from our statistical analysis, we estimate the lower- and upper-limit encounter frequencies for PSP. These two limits correspond to the minimum and maximum relative speed between the PSP footpoint and the solar surface. In the corotation case, the relative speed is taken to be approximately zero, so the effective scanning speed is minimized and PSP samples the smallest area, yielding the lower encounter frequency. In the perihelion case, the relative speed is maximal, so PSP sweeps across the largest effective area and yields the upper encounter frequency. Thus, the difference between the two values reflects the spacecraft sampling speed through Equation (5), rather than a change in the intrinsic occurrence rate of MFEs:

$$N'_{\rm PSP_mfe} \approx \begin{cases} 0.9\ {\rm day^{-1}}\ \text{(corotation)} \\ 6.05\ {\rm day^{-1}}\ \text{(perihelion)} \end{cases}. \tag{7}$$

Using PSP in situ measurements from the first five encounters, Y. Chen et al. (2021) applied a Grad–Shafranov reconstruction procedure, a single-spacecraft method that identifies intervals consistent with a two-dimensional magnetohydrostatic flux-rope structure, and identified 243 SMFRs over a total duration of 116 days, corresponding to a detection frequency of $\sim$2 events day$^{-1}$ at heliocentric distances of 0.13–0.66 au. In their classification, flux ropes represent relatively static flux-rope structures with weak Alfvénic signatures (Walén slope $<$0.3), whereas other flux ropes are accompanied by significant field-aligned flows (Walén slope $>$0.3 and correlation coefficient $R > 0.8$). We focus on the flux-rope population, as we expect that if MFEs inject coherent magnetic structures into the heliosphere, their in situ counterparts would resemble quasi-static flux ropes rather than predominantly Alfvénic intervals. Under this assumption, our estimated PSP encounter rate of MFE-related structures is consistent, to within an order of magnitude, with the flux-rope detection frequency reported by Chen et al. (2021). By contrast, chromospheric spicules, with occurrence rates of $\approx 10^6\,{\rm s^{-1}}$ over the Sun (e.g., J. Lee et al. 2024), are orders of magnitude more frequent and are believed to maintain the bulk mass and heat flux of the quiet solar wind. MFEs therefore occupy an intermediate regime: far less numerous than spicules but far more energetic, producing intermittent injections that can carry sufficient magnetic twist to form discrete transients—namely, SMFRs and localized field-line deflections such as switchbacks.

The enhanced MFE occurrence at CH boundaries, hence, provides a viable mechanism for intermittently perturbing the open-field corona. When successive MFEs erupt in these regions, the cumulative effect could bend or fold open-field lines, giving rise to the patchy switchback morphology measured by PSP and Solar Orbiter. Meanwhile, the subset of MFEs that fully reconnect with open flux may evolve into small, detached flux-rope structures convected outward as SMFRs. Thus, both phenomena can arise naturally from the same family of small-scale eruptions, differing primarily in reconnection geometry and coupling efficiency.

## 6. Summary

This study applied an automated detection and tracking algorithm for identifying MFEs using full-disk BBSO H$\alpha$ blue-wing data and analyzing them. This detection algorithm, which is based on intensity thresholding, feature labeling, temporal tracking, and eruption scoring, cataloged 1986 potential MFEs from nearly 4 hr of observation on 2020 June 9.

Four representative events were selected to illustrate the range of MFE behavior: a long-lived filament-like eruption (duration $\approx$140 minutes; length $\approx$100 Mm) showing coherent expansion; an intermediate event observed near a sunspot extending $\approx$40 Mm; another short, impulsive eruption lasts $\approx$10 minutes with peak size $\approx$30 Mm; and a mid-duration event observed, which reached $\approx$30 Mm. This study estimates a global occurrence rate of $6.61 \times 10^4$ MFEs per day.

Statistical analysis revealed that most eruptions are compact (mean size $\sim$17 Mm) and short-lived (mean lifetime $\sim$20 minutes), characteristic of minifilaments. As summarized in Table 1, eruptions near the AR had a slightly higher average length compared to the full-disk mean. This is consistent with magnetic topology in ARs supporting larger eruptions, though small-scale MFEs remain ubiquitous across the disk. The number density of MFEs within coronal holes is $\sim$20% lower than outside regions. Eruptions inside CHs are also shorter and smaller than those outside. Statistical tests confirm that both lifetime and length distributions differ significantly between regions, indicating that open, largely unipolar magnetic field environments suppress and constrain MFE activity.

A power-law behavior in the length distribution and a sublinear relationship between eruption length and duration were identified. Using MLE on the unbinned data—and applying an upper physical cutoff of 36.0 Mm to strictly isolate the minifilament regime—the main power-law index of the length distribution is calculated to be −4.78. Meanwhile, the empirical length–duration survival boundary follows length ∝ duration$^{0.22}$. While our derived length distribution slope of −4.78 is steeper than the −2.7 scaling index reported by A. C. Sterling & R. L. Moore (2016), the two characterize different statistical regimes. The −2.7 slope represents an interclass macroscaling envelope spanning multiple distinct phenomena (from CMEs to spicules), whereas our −4.78 slope describes the intraclass distribution strictly within the minifilament population. This steep internal decay suggests that while minifilaments are abundant enough to fit the global-scale-free continuum, local magnetic and geometric constraints cause the probability of supporting progressively larger coherent structures within this specific size regime to drop off rapidly. The comparison with previous studies using BBSO Hα data—N. Huang & H. Wang (2025), J. Wang et al. (2000), and L. M. Hermans & S. F. Martin (1986)—shows consistent occurrence rates once detection thresholds are adjusted, further supporting the robustness of this study's detection methodology.

In summary, our statistical results—combined with recent EUV and in situ studies—suggest that MFEs constitute a physically credible source of small-scale heliospheric transients. Their frequency and spatial preference near CH boundaries bridge the gap between abundant spicular outflows and rarer CMEs, providing a chromospheric-scale mechanism for structuring the solar wind across a wide range of scales.

## Acknowledgments

We gratefully acknowledge the use of data from the Big Bear Solar Observatory (BBSO), the Global Oscillation Network Group (GONG), and the Solar Dynamic Observatory Atmospheric Imaging Assembly (SDO/AIA). BBSO is operated by the New Jersey Institute of Technology and is supported by NSF grant AGS-2309939. GONG is operated by NISP/NSO/AURA/NSF with contributions from NOAA. This work was supported by NSF grants AGS-2114201, AGS-2229064, and AGS-2309939, and NASA grants 80NSSC19K0257, 80NSSC20K0025, 80NSSC20K1282, 80NSSC241914, and 80NSSC24K0258.

## ORCID iDs

Artin Khaleghi https://orcid.org/0009-0004-2330-6384
Qin Li https://orcid.org/0000-0002-3669-1830
Nengyi Huang https://orcid.org/0000-0001-9049-0653
Jeongwoo Lee https://orcid.org/0000-0002-5865-7924
Haimin Wang https://orcid.org/0000-0002-5233-565X